\documentclass[aps,prd,twocolumn,groupedaddress]{revtex4}

\usepackage{graphicx}
\usepackage{dcolumn}
\usepackage{bm}
\usepackage{amsmath}
\usepackage{epstopdf}
\usepackage{amsfonts}
\usepackage{amssymb}
\usepackage{epsfig}
\usepackage{tabularx}
\usepackage{color}

\usepackage[colorlinks=true,citecolor=blue,urlcolor=magenta,breaklinks]{hyperref}

\newcommand{\be}{\begin{equation}}
\newcommand{\en}{\end{equation}}
\newcommand{\bea}{\begin{eqnarray}}
\newcommand{\ena}{\end{eqnarray}}

\begin{document}

\title{Radial oscillations of strange quark stars\\
	admixed with condensed dark matter}

\author{G. Panotopoulos and  Il\'idio Lopes}
\email[]{grigorios.panotopoulos@tecnico.ulisboa.pt}
\email[]{ilidio.lopes@tecnico.ulisboa.pt}
\affiliation{Centro Multidisciplinar de Astrof\'{\i}sica, Instituto Superior T\'ecnico,
Universidade de Lisboa, Av. Rovisco Pais, 1049-001 Lisboa, Portugal}

\date{\today}

\begin{abstract}
We compute the 20 lowest frequency radial oscillation modes of strange stars admixed with condensed dark matter. We assume a self-interacting bosonic dark matter, and we model dark matter inside the star as a Bose-Einstein condensate. In this case the equation of state is a polytropic one with index $1+1/n=2$ and a constant $K$ that is computed in terms of the mass of the dark matter particle and the scattering length. Assuming a mass and a scattering length compatible with current observational bounds for self-interacting dark matter, we have integrated numerically first the Tolman-Oppenheimer-Volkoff equations for the hydrostatic equilibrium, and then the equations for the perturbations $\xi=\Delta r/r$ and $\eta=\Delta P/P$.
For a compact object with certain mass and radius we have considered here three cases, namely no dark matter at all and two different dark matter scenarios. Our results show that i) the separation between consecutive modes increases with the amount of dark matter, and ii) the effect is more pronounced for higher order modes. These effects are relevant even for a strange star made of 5\% dark matter.
\end{abstract}


\maketitle


\section{Introduction}

The concordance cosmological model based on cold dark matter and cosmological constant ($\Lambda$CDM) successfully describes the structure formation of the Universe from stars to galaxy clusters, as it is in excellent agreement with a vast amount of modern well-established astrophysical and cosmological observational data coming from many different sides. Thanks to the pioneer work of
F. Zwicky who studied the dynamical properties of the Coma galaxy cluster in 1933 \cite{zwicky}, and that of V. Rubin who studied the galaxy rotation curves in 1970 \cite{rubin}, it has been realized that 90\% of all matter in the present Universe is made of an unknown type of particles known as dark matter (DM). The determination of the type of elementary particles that play the role of dark matter in the Universe is one of the biggest challenges of particle physics and modern cosmology, since the origin and nature of DM still remain unknown, despite the fact that many DM candidates have been proposed and studied in the literature by cosmologists and particle physicists alike \cite{taoso,lopes} in order to constrain the properties of DM \cite{taoso,lopes}.
For a review on dark matter see e.g. \cite{munoz}.

Among all the possible DM candidates, perhaps the most popular class consists of the Weakly Interacting Massive Particles (WIMPs)
that are thermal relics from the Big-Bang, like photons and neutrinos.
Initially in the early hot Universe the temperature was high enough to maintain the DM particles $\chi$ in equilibrium with the particles of the Standard Model. However, at a certain point as the Universe expands and cools down, the DM particles can no longer annihilate, they decouple from the thermal bath, and consequently their abundance freezes-out and remains constant ever since.
Accordingly, today's DM relic density is given $\Omega_{\chi} h^2 = {3 \times 10^{-27} cm^3 s^{-1}}/{\langle \sigma v \rangle_\chi}$ \cite{SUSYDM}, where $h$ is related to the Hubble constant $H_0=100 \: h (km s^{-1})/(Mpc)$. The numerical value of the typical WIMP annihilation cross section $\langle \sigma v \rangle_\chi = 3 \times 10^{-26} cm^3/s$, characterizing weak interactions, is capable of reproducing the current observed DM abundance $\Omega_\chi h^2=0.1198 \pm 0.0015$ \cite{wmap,planck2015}.

In the standard picture WIMPs as described in the previous paragraph are assumed to be collisionless. This, however, is just one possibility among others. As a matter of fact long time ago in the literature it was introduced the idea that the dark matter particles may have self-interactions \cite{self-interacting}. In this newly proposed scenario some apparent conflicts between astrophysical observations from the one hand and the collisionless cold dark mater paradigm from the other could be eliminated or at least alleviated \cite{self-interacting}. It was found in that work that the DM self-interaction cross section has to lie in the following range $ 0.45\; cm^2g^{-1} < \sigma_\chi/m_\chi < 450 \;cm^2g^{-1} $ where $m_\chi$ is the mass of the dark matter particles, and $\sigma_\chi$ the self-interaction cross section of dark matter. Nowadays current limits on the strength of the dark matter self-interaction require that the ratio $\sigma_\chi/m_\chi$ satisfies the bounds $1.75 \times 10^{-4} \;cm^2g^{-1} < \sigma_\chi/m_\chi < (1-2) \;cm^2g^{-1} $ \cite{bullet1,bullet2,review}.

Compact objects, such as neutron stars or strange stars \cite{textbook,strangestars}, comprise excellent natural laboratories to study, test and constrain new physics and/or alternative theories of gravity under extreme conditions that cannot be reached to earth-based experiments. It is well-known that the properties of compact objects, such mass and radius, depend crucially on the equation of state which unfortunately is poorly known. Studying the oscillations of stars and computing the frequency modes offer us the opportunity to probe the interior of the stars and learn more about the equation of state, since the precise values of the frequency modes are very sensitive to thermodynamics of the internal structure of the star. This work is substantiated by the fact that there is some evidence that strange stars may exist. Indeed the recent discovery of very  compact objects (with very high densities)  like  the  millisecond pulsars SAX J 1808.4-3658 and RXJ185635-3754, the X-ray burster 4U 1820-30, the X-ray pulsar Her X-1, and  X-ray source PSR 0943+10 are among the best candidates~\citep{2012ARNPS..62..485L}.
Moreover, the recently launched NASA mission NICER, designed primarily to observe thermal X-rays emitted by several millisecond pulsars, could help answer to this question~\citep{2014SPIE.9144E..20A}.

To the best of our knowledge not much work has been done on the study of radial oscillations of compact objects admixed with dark matter. Previously the authors of \cite{ref1,ref2} studied the equilibrium configuration and the radial oscillation modes of DM admixed neutron stars, where DM was modeled as an ideal gas of non-interacting fermions. It is the aim of the present article to study the effects of bosonic condensed dark matter on strange stars as far as the radial oscillations are concerned. Our work is organized as follows: after this introduction, we present the equation of state
in section two, and we present the equations for the perturbations in the the third section. Our numerical results are discussed
in section four, and finally we conclude in the last section. We shall be working in natural units in which the speed of light in vacuum $c$ as well as the reduced Planck constant $\hbar$ are set to unity, $c=1=\hbar$.
In these units all dimensionful quantities are measured in GeV, and we shall making use of the following conversion rules
$1 m = 5.068 \times 10^{15} GeV^{-1}$ and $1 kg = 5.610 \times 10^{26} GeV$ \cite{guth}.

\section{Equation of state}

For strange matter we shall consider the simplest equation of state corresponding to a relativistic gas of de-confined quarks,
known also as the MIT bag model \cite{bagmodel}
\be
P_s = \frac{1}{3} (\epsilon_s - 4B)
\en
and the bag constant has been taken to be $B=(148 MeV)^4$ \cite{Bvalue}.
We mention in passing that for refinements of the bag model the reader could consult e.g. \cite{refine}, and for the present state-of-the-art the recent paper \cite{art}. Strange stars with the above equations of state have been
investigated in \cite{prototype} for negative and positive values of the cosmological constant. There it was 
shown that the observed value of the cosmological constant $\Lambda \sim (10^{-33} eV)^2$ is too small to have an observable effect. That is why in the present work we have taken the cosmological constant to be zero.

For the condensed dark matter we shall consider the equation of state obtained in \cite{darkstars}, namely $P=K \epsilon^2$,
where the constant $K=2 \pi l/m_\chi^3$ is given in terms of the mass of the dark matter particles $m_\chi$ and the scattering length $l$.
In a dilute and cold gas only the binary collisions at low energy are relevant, and these collisions are characterized by the s-wave scattering length
$l$ independently of the form of the two-body potential \cite{darkstars}. Therefore we can consider a short range repulsive delta-potential of the form
\begin{equation}
V(\vec{r}_1-\vec{r}_2) = \frac{4 \pi l}{m_\chi} \delta^{(3)}(\vec{r}_1-\vec{r}_2)
\end{equation}
which implies a dark matter self interaction cross section of the form $\sigma_\chi=4 \pi l^2$ \cite{darkstars,chinos}.
Assuming a mass $m_\chi = 0.32~GeV$ and a scattering length $l=2.12~fm$, or a mass $m_\chi = 0.44~GeV$ and a scattering length $l=2.49~fm$
the bounds discussed in the Introduction are satisfied while the constant $K$ is found to be
\begin{equation}
K = \frac{1.01}{B}
\end{equation}
for the first case (hereafter first DM model), and
\begin{equation}
K = \frac{0.46}{B}
\end{equation}
for the second case (hereafter second DM model).

\section{Radial oscillations}

In this section we present the basic differential equations to be integrated numerically, namely the TOV equations for the interior solution, and then the equations for the perturbations, which will allow us to compute the oscillation modes.

\subsection{Hydrostatic equilibrium}

We briefly review the standard equations for 
relativistic stars in General Relativity (GR). Accordingly,
Einstein's field equations without a cosmological constant read
\be
G_{\mu \nu} = R_{\mu \nu}-\frac{1}{2} R g_{\mu \nu}  = 8 \pi T_{\mu \nu}
\en
where we have set Newton's constant $G$ equal to unity, and in the exterior problem the matter energy momentum tensor $T_{\mu \nu}$ vanishes.
Moreover, for matter we assume a perfect fluid with an energy density $\epsilon$, a pressure $P$, with an equation of state  $P(\epsilon)$.
As usual in static spherically symmetric spacetimes we 
we consider for the metric the ansatz
\be
ds^2 = -f(r) dt^2 + g(r) dr^2 + r^2 d \Omega^2
\en
or
\be
ds^2 = -e^\nu(r) dt^2 + \frac{1}{1-2 m(r)/r} dr^2 + r^2 d \Omega^2
\en
with two unknown metric functions of the radial distance $f(r), g(r)$ or $\nu(r), m(r)$. For the exterior problem $r > R$, with $R$ being the
radius of the star, one obtains the well-known solution
\be
f(r) = g(r)^{-1} = 1-\frac{2 M}{r}
\en
where $M$ is the mass of the star.
For the interior solution $r < R$ we use the functions $\nu(r), m(r)$ instead of the functions $f(r), g(r)$
so that upon matching the two solutions at the surface of the star we obtain $m(R)=M$.
The Tolman-Oppenheimer-Volkoff (TOV) equations for the interior solution of a relativistic star with a
vanishing cosmological constant read \cite{TOV}
\bea
m'(r) & = & 4 \pi r^2 \epsilon(r) \\
P'(r) & = & - (P(r)+\epsilon(r)) \: \frac{m(r)+4 \pi P(r) r^3}{r^2 (1-\frac{2 m(r)}{r})} \\
\nu'(r) & = & -2 \frac{P'(r)}{\epsilon(r)+P(r)}
\ena
where the prime denotes differentiation with respect to r. The first two equations are to be integrated with the initial conditions
$m(r=0)=0$ and $P(r=0)=P_c$, where $P_c$ is
the central pressure. The radius of the star is determined requiring that the energy density vanishes at the surface,
$P(R) = 0$, and the mass of the star is then given by $M=m(R)$. Finally, the other metric function can be computed using the third equation
together with the boundary condition $\nu(R)=ln(1-2M/R)$.

Now let us assume that the star consists of two fluids, namely strange matter (de-confined quarks) and dark matter with only gravitational interaction between
them, and equations of state $P_s(\epsilon_s)$, $P_\chi(\epsilon_\chi)$ respectively. The total pressure and the total energy density of the system are given by $P=P_s+P_\chi$ and $\epsilon=\epsilon_s+\epsilon_\chi$ respectively.
Since the energy momentum tensor of each fluid is separately conserved, the TOV equations in the two-fluid formalism for the interior solution
of a relativistic star with a vanishing cosmological constant read \cite{2fluid1,2fluid2}
\bea
m'(r) & = & 4 \pi r^2 \epsilon(r) \\
P_s'(r) & = & - (P_s(r)+\epsilon_s(r)) \: \frac{m(r)+4 \pi P(r) r^3}{r^2 (1-\frac{2 m(r)}{r})} \\
P_\chi'(r) & = & - (P_\chi(r)+\epsilon_\chi(r)) \: \frac{m(r)+4 \pi P(r) r^3}{r^2 (1-\frac{2 m(r)}{r})}
\ena
In this case in order to integrate the TOV equations we need to specify the central values both for normal matter and for
dark matter $P_s(0)$ and $P_\chi(0)$ respectively. So first we define the dark matter fraction
\be
f = \frac{P_\chi(0)}{P_s(0)+P_\chi(0)}
\en
and we consider two cases, namely $f=0.09, 0.2$. We have chosen these values in agreement with the current dark matter constraints
obtained from stars like the Sun. Actually, as shown by several authors, even smaller amounts of DM (as a percentage of the total mass of the star)
can have a quite visible impact on the structure of these stars \cite{silk,ilidio2,ilidio5}.

In the following we will compute the frequency radial oscillation modes for the no DM case as well as for two DM models
shown below

\begin{equation}
\textrm{First DM model} \rightarrow
\left\{
\begin{array}{lcl}
K = \frac{1.01}{B} \\
&
&
\\
 f = 0.09
\end{array}
\right.
\end{equation}

corresponding to a DM mass fraction $M_{dm}/M \simeq 5 \%$

and

\begin{equation}
\textrm{Second DM model} \rightarrow
\left\{
\begin{array}{lcl}
K = \frac{0.46}{B} \\
&
&
\\
 f = 0.20
\end{array}
\right.
\end{equation}

corresponding to a DM mass fraction $M_{dm}/M \simeq 12 \%$. All three cases considered here predict a compact star with the following properties

\begin{equation}
\textrm{Properties of compact star} \rightarrow
\left\{
\begin{array}{lcl}
\textrm{star mass} \; M = 1.77 \: M_{\odot} \\
&
&
\\
\textrm{star radius} \; R = 10.94 \: km
\end{array}
\right.
\end{equation}

As we discuss in this work, even such small amounts of DM can change the frequency radial oscillation modes of strange stars significantly.
\begin{figure}[ht!]
	\centering
	\includegraphics[scale=0.7]{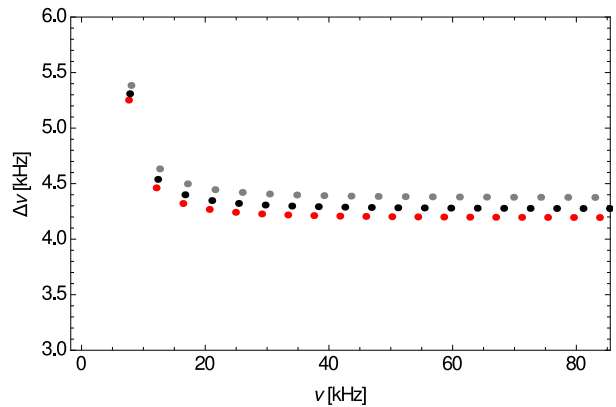}
	\caption{Frequency differences $\Delta\nu_{n}=\nu_{n+1}-\nu_n$ versus $\nu_n$ in kHz for a) no DM (red points), first DM model (black points) and second
		DM model (gray points).}
	\label{fig:Freq} 	
\end{figure}

\subsection{Equations for the perturbations}

If $\Delta r$ is the radial displacement and $\Delta P$ is the perturbation of the pressure, the equations governing the dimensionless
quantities $\xi=\Delta r/r$ and $\eta=\Delta P/P$ are the following \cite{chanmugan1,chanmugan2}
\be
\xi'(r) = -\frac{1}{r} \left( 3 \xi + \frac{\eta}{\gamma} \right) - \frac{P'(r)}{P+\epsilon} \xi(r)
\en
\begin{widetext}
\be
\begin{split}
\eta'(r) = \xi \left[ \omega^2 r (1+\epsilon/P) e^{\lambda-\nu} - \frac{4 P'(r)}{P} - 8 \pi (P+\epsilon) r e^{\lambda} + \frac{r (P'(r))^2}{P (P+\epsilon)}\right]
+ \eta \left[ -\frac{\epsilon P'(r)}{P (P+\epsilon)}-4 \pi (P+\epsilon) r e^{\lambda} \right]
\end{split}
\en
\end{widetext}
where $e^{\lambda}, e^{\nu}$ are the two metric functions, $\omega$ is the frequency oscillation mode, and $\gamma$ is the relativistic
adiabatic index defined to be
\begin{equation}
\gamma = \frac{d P}{d \epsilon} (1+\epsilon/P)
\end{equation}
Note that contrary to \cite{ref2} where the authors considered two families of modes, namely one related to perturbations of the ordinary matter and another
related to the perturbations of the DM fluid, in this work we have considered the total perturbation of the pressure.
The system of two coupled first order differential equations is supplemented with two boundary conditions, one at the center as $r \rightarrow 0$, and
another at the surface $r=R$.
The boundary conditions are obtained as follows: In the first equation, $\xi'(r)$ must be finite as $r \rightarrow 0$, and therefore we require
that
\be
\eta = -3 \gamma \xi
\en
must satisfied at the center. Moreover, in the second equation, $\eta'(r)$ must be finite at the surface as $\epsilon,P \rightarrow 0$ and therefore
we demand that
\be
\eta = \xi \left[ -4 + (1-2M/R)^{-1}  \left( -\frac{M}{R}-\frac{\omega^2 R^3}{M} \right )  \right]
\en
must satisfied at the surface, where we recall that $M,R$ are the mass and the radius of the star respectively. Using the shooting method we first compute
the dimensionless quantity $\bar{\omega} = \omega t_0$ where $t_0=1 ms$. Then the frequencies are computed by
\be
\nu = \frac{\bar{\omega}}{2 \pi} \; kHz
\en
Therefore, contrary to the previous hydrostatic equilibrium problem, which is an initial value problem,
this is a Sturm-Liouville boundary value problem,
and as such the frequency $\nu$ is allowed to take only particular values, the so-called eigenfrequencies $\nu_n$. Each $\nu_n$ corresponds to a specific radial oscillation mode of the star. Accordingly, each radial mode is identified by its $\nu_n$ and  by an associated pair of eigenfunctions -- the displacement perturbation $\xi_n(r)$  and the pressure perturbation $\eta_n(r)$. In simple stellar models as the one considered in this study, the order of the mode $n$ is equal to the number of nodes of the radial eigenfunction  $\xi_n(r)$~\citep[e.g.,][]{2000ApJ...542.1071L,1989nos..book.....U}.

\begin{figure}[ht!]
\centering
\includegraphics[scale=0.6]{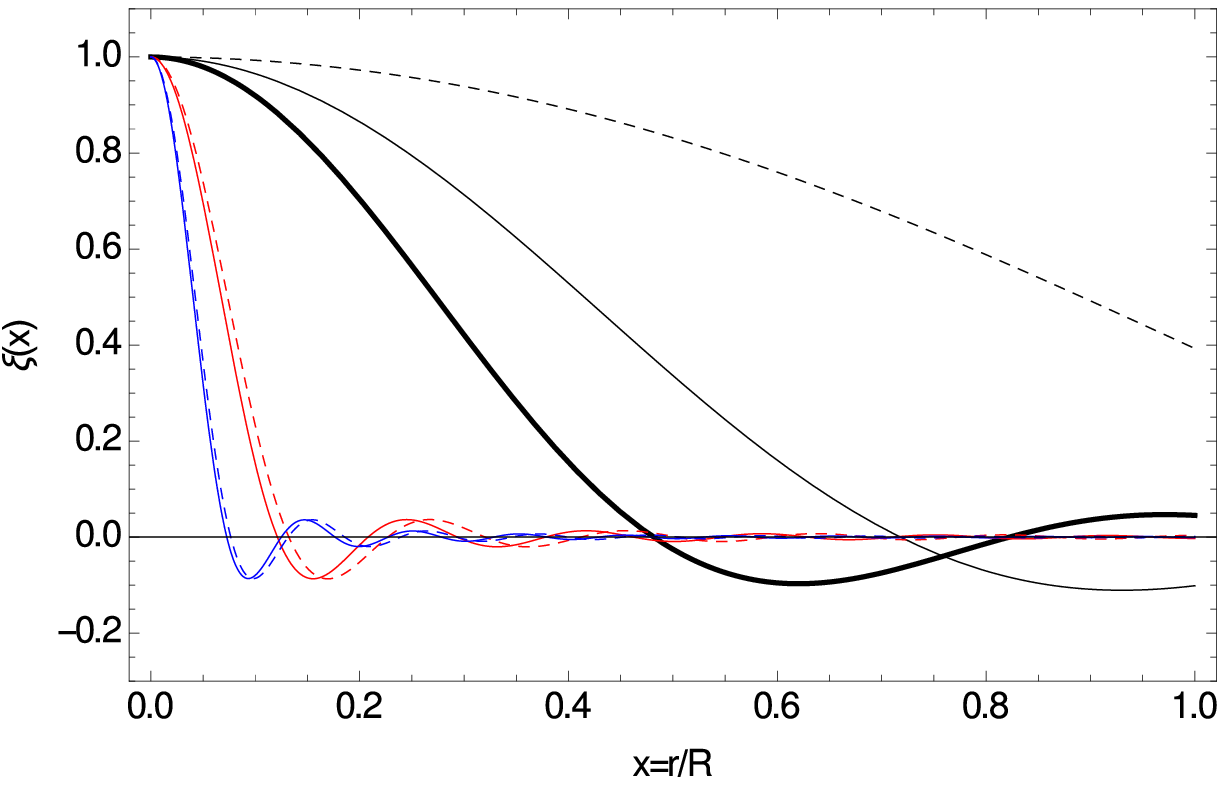}
\includegraphics[scale=0.6]{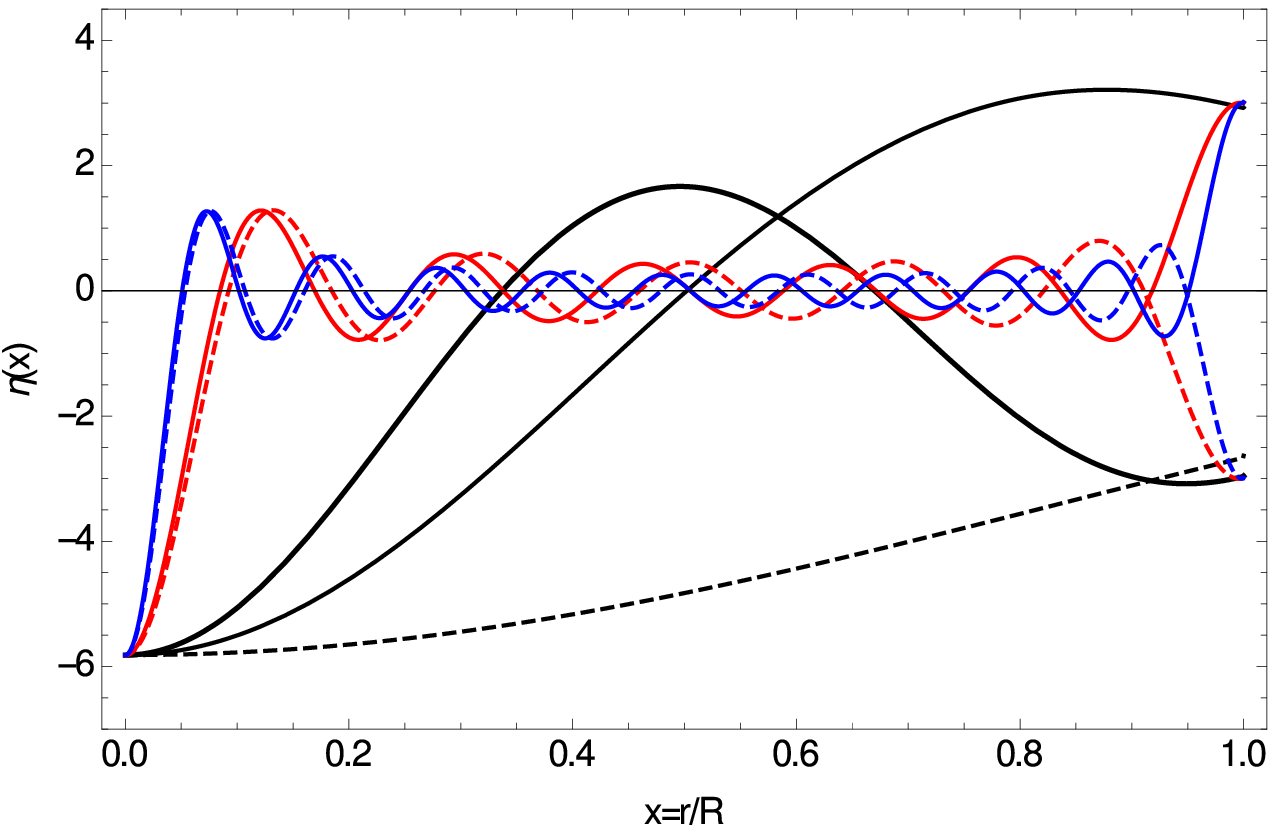}
\caption{Perturbations $\xi_n(r)$ and  $\eta_n(r)$ versus $x=r/R$ for low order modes $n=0,1,2$ (in black), for intermediate modes $n=10,11$ (in red), and highly excited modes $n=18,19$ (in blue) for the no DM case.}
\label{fig:Eta} 	
\end{figure}

\section{Numerical results}

As we have already mentioned, in this work we have considered a fiducial compact star with a mass $M=1.77 M_{\odot}$ and radius $R=10.94 km$ in 3 cases,  namely a) no dark matter at all, b) a strange star made of 5 per cent of dark matter, and c) a strange star made of 12 per cent of dark matter.
The results are summarized in the table below. We have computed the first 20 frequencies, however in the table~\ref{table:Firstset} we show some of them, while the rest
can be seen in Fig.~\ref{fig:Freq}. We notice that the separation between consecutive modes increases with the amount of dark matter, and
moreover the effect is more pronounced for higher order modes. In \cite{ref2}, although the physics of the model is different, the authors found for the first four oscillation modes frequency values of the same order of magnitude. In addition, they found that the frequencies for the higher order modes increase with the DM mass fraction, in agreement with our results.

We recall that in a Sturm-Liouville boundary value problem the number of zeros of the eigenfunctions corresponds to the overtone number $n$, namely the first excited mode corresponds to $n=1$ and has only one zero, the second excited mode corresponds to $n=2$ and has two zeros, while the fundamental mode does not have zeros and corresponds to $n=0$. The fundamental mode ($n=0$) is also known as the f-mode and the
other modes with $n\ne0$ are the so-called p-modes (pressure modes or acoustic modes)~\citep[e.g.,][]{1989nos..book.....U}.
In Figure~\ref{fig:Eta} we show three groups of eigenfunctions, namely low modes $n=0$ (f-mode) and $n=1,2$ shown in black, intermediate modes $n=10,11$ shown in red, and finally highly excited modes ($n=18,19$) shown in blue. In particular, the fundamental as well as the
$n=10$ and the $n=18$ modes are represented by dashed curves.

This shows that if radial oscillations are discovered in
compact stars, specifically in strange stars, it will be possible to use
their frequencies to infer the properties of their internal structure,
and check if there is evidence for dark matter inside the star.
As in any normal star, the  easiest and most simple diagnostic
of interest to observers will be to  compute differences between consecutive
modes, the so-called large separation~\citep[e.g.,][]{2001A&A...373..916L,1989nos..book.....U} here shown in Figure~\ref{fig:Freq}. The amount of DM present inside the star, even in small quantities ($\sim 5 $ -- $10\%$) has a strong impact on
the large separation $\Delta\nu_{n}$ which, depending on the order
of the mode $n$, is sensitive to different regions of the star.

\begin{table}
\begin{tabular}{l | l l l}
mode & \multicolumn{3}{c}{{\sc Compact star's models}}\\
order $n$ & No DM & $1^{st}$ DM & $2^{nd}$ DM \\
\hline
\hline
0  &  2.35  & 2.47  & 2.60  \\
1 & 7.60 & 7.77 & 7.99  \\
2 & 12.06 & 12.31 & 12.62 \\
3 & 16.38 & 16.71 & 17.12  \\
4 & 20.65 & 21.06 & 21.57 \\
5 & 24.89 & 25.38 & 25.99 \\
$\cdots$ & $\cdots$ & $\cdots$ & $\cdots$ \\
10 & 45.96 & 46.85 & 47.96 \\
11 & 50.17 & 51.14 & 52.34  \\
$\cdots$ & $\cdots$ & $\cdots$ & $\cdots$ \\
18 & 79.55 & 81.08 & 82.99 \\
19 & 83.74 & 85.36 & 87.36
\end{tabular}
\caption{Frequencies $\nu_n$ in $kHz$ for all 3 compact
	stellar models	considered here (see Figures~\ref{fig:Freq} and~\ref{fig:Eta}).}
\centering
\label{table:Firstset}
\end{table}

This result can be interpreted as follows: Figure~\ref{fig:Eta} shows the perturbations $\xi_n(r)$  and  $\eta_n(r)$ as a function of the dimensionless radial distance $x=r/R$ (for the no DM case) for the  fundamental mode ($n=0$) as well as for several excited modes. The amplitude of $\xi_n(r)$ for each mode of frequency $\nu_n$ is larger closer to center (but not the center) and much smaller near the surface. Alternatively, the amplitude of $\eta_n(r)$  is larger closer to the center and near the surface of the star. Hence, it results that $\xi_{n+1}(r)-\xi_n(r)$ and $\eta_{n+1}(r)-\eta_n(r)$ are more sensitive to the star's core. As such $\Delta\nu_{n}$ ($=\nu_{n+1}-\nu_n$) is an observational imprint of the deepest layers of the interior of this star.  Notice that although  $\eta_n(r)$ of consecutive $n$ have large amplitudes near the surface with opposite signal (opposition of phase) its contribution for $\eta_{n+1}(r)-\eta_n(r)$ cancels  out.  Therefore,
the $\Delta\nu_{n}$  is a convenient method to measure
the content of DM inside strange stars, specifically in its central region.


\section{Conclusions}

In this work we have studied the radial oscillations of strange stars admixed with condensed dark matter, and we have computed numerically the 20 lowest
frequency radial oscillation modes. We assume a self-interacting bosonic dark matter, and
we model dark matter inside the star as a Bose-Einstein condensate. In this case the equation of state is found to be $P=K \epsilon^2$ where
the constant $K=2 \pi l/m_\chi^3$ is computed in terms of the mass of the dark matter particle $m$ and the scattering length $l$.
Assuming a mass and a scattering length compatible with current observational bounds for self-interacting dark matter, we have integrated
numerically first the TOV equations for the hydrostatic equilibrium in the two-fluid formalism, and then the equations for the perturbations
$\xi=\Delta r/r$ and $\eta=\Delta P/P$. Our results are summarized in Table~\ref{table:Firstset} and in Figures~\ref{fig:Freq} and~\ref{fig:Eta}. For a compact object with certain mass and radius we have considered here three cases, namely no dark matter at all and two different dark matter scenarios. Our results show that i) the  separation between consecutive modes (large separation) increases with the amount of dark matter, and ii) the effect is more pronounced for higher order modes.
Thus, the high sensitivity of the large separation to the internal structure of the star can be used to probe the DM content present inside the strange star.


\section*{Acknowlegements}

The authors thank the Funda\c c\~ao para a Ci\^encia e Tecnologia (FCT), Portugal, for the financial support to the Multidisciplinary
Center for Astrophysics (CENTRA),  Instituto Superior T\'ecnico,  Universidade de Lisboa,  through the Grant No. UID/FIS/00099/2013.
We wish to thank the anonymous reviewer for valuable comments and suggestions.



\end{document}